\documentclass[twocolumn,english,showpacs,floatfix,amsmath,amssymb,prl,superscriptaddress]{revtex4}
\usepackage[T1]{fontenc}
\usepackage[latin9]{inputenc}
\usepackage{babel}

\usepackage{graphicx}
\usepackage{amssymb}
\usepackage{esint}



\begin{document}

\title{Energy relaxation and thermalization of hot electrons in quantum wires}

\author{Torsten Karzig} 
\affiliation{\mbox{Dahlem Center for Complex Quantum Systems and Fachbereich Physik, Freie Universit\"at Berlin, 14195 Berlin, Germany}}

\author{Leonid I. Glazman} 
\affiliation{Department of Physics, Yale University, 217 Prospect Street,
New Haven, Connecticut 06520, USA}

\author{Felix von Oppen}
\affiliation{\mbox{Dahlem Center for Complex Quantum Systems and Fachbereich Physik, Freie Universit\"at Berlin, 14195 Berlin, Germany}}

\date{\today}
\begin{abstract}
We develop a theory of energy relaxation and thermalization of hot carriers in real quantum wires. Our theory is based on a controlled perturbative approach for large excitation energies and emphasizes the important roles of the electron spin and finite temperature. Unlike in higher dimensions, relaxation in one-dimensional electron liquids requires three-body collisions and is much faster for particles than holes which relax at nonzero temperatures only. Moreover, co-moving carriers thermalize more rapidly than counterpropagating carriers. Our results are quantitatively consistent with a recent experiment.
\end{abstract}
\pacs{71.10.Pm}
\maketitle

{\em Introduction.---}The behavior of electrons confined to move in one spatial dimension is frequently described within the Tomonaga-Luttinger model which assumes a linear dispersion relation for the electrons. In this model, all excitations move at the same velocity so that electron-electron interactions become particularly significant. Consequently, the electron system can no longer be described as a Fermi liquid but instead, is expected to form a Luttinger liquid. In recent years, much effort has been expended on elucidating the consequences of Luttinger-liquid physics in quantum wires \cite{Review10}. 

A peculiar consequence of the Tomonaga-Luttinger model is the complete absence of inelastic processes for hot particles or holes. As emphasized by a recent experiment \cite{Barak10}, the physics of energy relaxation is much richer in real quantum wires with a nonlinear dispersion. In this experiment, hot carriers of well-defined energy and momentum are injected into a quantum wire and their energy relaxation is probed in cleverly designed transport measurements. The experiment shows not only that hot carriers relax but also that energy relaxation is much more efficient for hot particles than for hot holes, in stark contrast to electron liquids in higher dimensions. Moreover, a simple model \cite{Barak10} reproducing the experimental observations assumed that thermalization occurs much faster among co-moving electrons than between right- and left-moving carriers. 

Foci of recent theoretical work on one-dimensional electron systems were nonequilibrium effects \cite{Mirlin08,Takei09} and consequences of a nonlinear dispersion \cite{Khodas07,Teber07,Pereira09,Imambekov09, Pereira10, Essler10}. Nonequilibrium physics of systems with nonlinear dispersions has been accessible within a perturbative approach for weak interactions \cite{Lunde07,Rech08,Micklitz10}. The latter is peculiar because pair collisions are ineffective for a quadratic dispersion. Indeed, by momentum and energy conservation, pair collisions result either in zero-momentum transfer or exchange of the momenta of the colliding particles. Both processes do not change the electronic distribution function. A kinetic theory of real one-dimensional electron systems therefore involves three-body collisions \cite{Lunde07}. 

Consider an electron injected into the quantum wire with an excitation energy $\epsilon$ above the Fermi energy $\epsilon_{F}$. Due to the quadratic dispersion $\epsilon_k=\hbar^2k^2/2m$, its velocity differs from that of the electrons in the Fermi sea by at least $\Delta v=\epsilon/mv_F$. (Here, $v_{F}$ is the Fermi velocity.) According to the standard condition for the validity of the Born approximation in quantum mechanics \cite{Landau_QM}, we therefore expect a perturbative approach to energy relaxation and thermalization to be appropriate when $\epsilon \gg mv_F\tilde U(0)/\hbar$, with $\tilde U(q)$ denoting the Fourier transform of the electron-electron interaction. From the point of view of Luttinger-liquid theory, this condition ensures that the electrons retain their integrity during the collision process. Indeed, at weak coupling the difference between the spin and charge velocities of the Luttinger liquid is $v_c-v_s\simeq \tilde U(0)/\pi\hbar$, so that the condition can be recast as $v_c-v_s\ll \Delta v$. Hence, spin and charge do not separate appreciably during the collision process. 

It is the purpose of the present paper to develop a theory of energy relaxation and thermalization in quantum wires in this perturbative regime. While some of the basic physics -- such as the asymmetry between hot particles and holes \cite{Nozieres94,Khodas07, Pereira09} -- was already understood in previous work, no systematic and quantitative theory of this fundamental property of one-dimensional electron systems exists to date. Specifically, we emphasize the important roles played by finite temperature and spin which are not appreciated in the existing literature and important for understanding the experiment \cite{Barak10}.

\begin{figure*}[t]
\includegraphics[width=14cm, keepaspectratio=true]{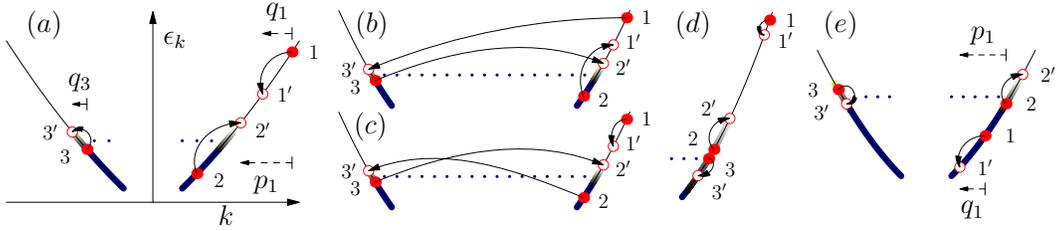} 
\caption{(Color online) Basic three-body relaxation processes of hot particles (a-d) and hot holes (e): (a) Small-$q$ process $T^{123}_{1'2'3'}$ and $2k_F$-processes (b) $T^{123}_{3'1'2'}$ and (c) $T^{123}_{1'3'2'}$. The remaining processes follow by exchanging $1'\leftrightarrow 2'$. (d) Competing relaxation process involving only co-moving electrons. (e) A small-$q$ relaxation process of hot holes. Dotted lines: $\epsilon_F$. 
\label{fig:processes}}
\end{figure*}

{\em Basic processes and results.---}The asymmetry in energy relaxation between hot particles and holes can be readily understood from the basic three-body collisions as sketched in Fig.\ \ref{fig:processes}(a)-(c),(e). Suppose a hot particle $1$ on, say, the right-moving branch transfers momentum $q_1$ to a right-mover in the Fermi sea. Due to the positive curvature of the dispersion relation, the energy loss $\Delta\epsilon$ of the hot particle exceeds the energy of the created particle-hole pair. This mismatch can be fixed by simultaneously exciting a left-moving particle-hole pair [Fig.\ \ref{fig:processes}(a)]. In line with the energy mismatch, the energy transfer to the left-moving particle-hole pair is of order $(\epsilon/ \epsilon_F) \Delta \epsilon$. The typical energy loss $\Delta\epsilon$ of hot particles in a single three-body collision is of order $\epsilon$.

Compare this with the relaxation of hot holes sketched in Fig.\ \ref{fig:processes}(e) where, for a given momentum transfer, the energy gain due to filling the hole by a higher-energy electron is smaller than the energy cost of exciting a co-moving particle-hole pair. Fixing this energy mismatch is therefore possible only if a counterpropagating electron gives up energy. Clearly, the Pauli principle forbids such a process at zero temperature and the hole is unable to relax. Indeed, this conclusion remains true for arbitrary $n$-body processes \cite{Barak10}. 

Hot holes do relax, however, at finite temperatures. Due to thermal smearing, the counterpropagating electron can give up an energy of order $T$. Thus, the hot hole can relax its energy, with a maximal energy loss of $\Delta\epsilon \sim \epsilon_FT/\epsilon$. This implies that hot holes float towards the Fermi energy in many small steps as long as $\epsilon \gg \epsilon_T$. Here, $\epsilon_T = \sqrt{\epsilon_F T}$ is a  new characteristic energy scale introduced by finite temperature. Conversely, for $\epsilon\ll\epsilon_T$, the energy loss of the hole per three-body collision would be comparable to its energy.  

Although hot particles always relax in a small number of three-body collisions, the energy scale $\epsilon_T$ is also relevant in this case. Indeed, for $\epsilon\ll\epsilon_T$, the energy transfer to the counterpropagating particle-hole pair is small compared to temperature. This has two important consequences. First, the phase space of the left-moving particle-hole pair is no longer controlled by the typical energy transfer $\epsilon^2/\epsilon_F$ but by temperature $T$ so that the energy relaxation rate of hot particles becomes temperature dependent. Second, it is no longer relevant whether the counterpropagating electron gains or loses energy so that for $\epsilon \ll \epsilon_T$, energy relaxation becomes equally fast for hot particles and holes.

Although these three-body collisions generate the same number of right- and left-moving particle-hole pairs, there is a fundamental asymmetry due to the parametric difference in their energies. Due to the larger energy transfer to the co-moving particle-hole pair, thermalization will happen more rapidly between electrons of the same chirality than between electrons of opposite chiralities. Injection of, say, right-moving hot particles can thus lead to different temperatures of right- and left-moving electrons over significant distances as appears to be the case in experiment \cite{Barak10}. We also note that the relevant thermalization rate among electrons of the same chirality is controlled by the hole relaxation rate even when injecting hot particles, since the relaxation of high-energy particles necessarily involves the excitation of deep holes.

We now summarize our quantitative results for the various energy relaxation rates of a spinful electron liquid interacting via the Coulomb interaction. (The derivation of these results is sketched further below.) For the energy relaxation rate of hot particles, we find   
\begin{equation}
{1}/{\tau_{{\rm p}}}=({9\epsilon_{F}}/{32\pi^{3}\hbar}) ({e^{2}}/{\kappa\hbar v_{F}})^{4}[\lambda(\epsilon)]^2({\epsilon}/{\epsilon_{F}})^{2}
\label{ElRelRate}
\end{equation}
at high energies $\epsilon\gg\epsilon_T$, and
\begin{equation}
{1}/{\tau_{{\rm p}}}=({3c_1\epsilon_{F}}/{4\pi^3\hbar})({e^{2}}/{\kappa\hbar v_{F}})^{4}[\lambda(\epsilon)]^2({T}/{\epsilon_{F}})
\label{EHRel}
\end{equation}
at low energies $\epsilon\ll\epsilon_T$. Here, $\lambda(\epsilon)=\ln|1/2k_{{\rm F}}a|\ln|\epsilon/4\epsilon_{{\rm F}}|$ is a dimensionless coupling constant in terms of the wire width $a$ and the Fermi wavevector $k_F$ with $k_Fa\ll 1$, $\kappa$ the dielectric constant of the host material and $c_1=4\ln(2)-1$. 

For $\epsilon\gg\epsilon_T$, the hole relaxation rate 
\begin{equation}
{1}/{\tau_{{\rm h}}}=({2\epsilon_{{\rm F}}}/{\pi\hbar})({e^{2}}/{\kappa\hbar v_{F}})^{4}[\lambda(\epsilon)]^2({T/\epsilon})^{2}.
\label{HoRel}
\end{equation}
is smaller than the particle relaxation rate (at a comparable excitation energy) by a factor $(\epsilon_T/\epsilon)^4$.  In the opposite limit, $\epsilon\ll\epsilon_T$, one may use Eq.\ (\ref{EHRel}) also for hole relaxation. Finally, a small temperature difference between right- and left-moving carriers equilibrates at the rate
\begin{equation}
{1}/{\tau_{{\rm inter}}}=(9 c_2 {\epsilon_{F}}/{2^8 \pi^5\hbar})({e^{2}}/{\kappa\hbar v_{F}})^{4}[\lambda(T)]^2({T}/{\epsilon_{{\rm F}}})^{3},
\label{TRel}
\end{equation}
with $c_2\approx103.9$ \cite{integral}. The results in Eqs.\ (\ref{ElRelRate})-(\ref{TRel}), based on the processes shown in Fig.\ \ref{fig:processes}(a)-(c),(e), dominate over the competing class of processes of Fig.\ \ref{fig:processes}(d) if at least one of two conditions is met: temperature is not too high, $T\ll\epsilon_F/(k_Fd)^4$, or energy is not too low, $\epsilon\gg\epsilon_T$. (Here, $d$ is the distance to a nearby metallic gate.)

{\em Evaluation of the energy relaxation rates.---}Our theory is based on the Boltzmann equation for the electronic distribution function $n(k,t)=n^{0}(k)+\delta n(k,t)$, linearized about the equilibrium Fermi-Dirac distribution $n^{0}(k)$,
\begin{equation}
\partial_{t}n_{1}  =  -\sum_{231'2'3'}W^{123}_{1'2'3'}n_{1}^{0}n_{2}^{0}n_{3}^{0}\bar n_{1'}^{0}\bar n_{2'}^{0}\bar n_{3'}^{0}\sum_{i=1}^3[y_i-y_{i'}].
\label{eq:boltzmann}
\end{equation}
Here, $y_i=\delta n_i/[n_{i}^{0}\bar n_{i}^{0}]$, $n_{i}=n(k_{i})$, and $\bar n_i=1-n_i$. The initial (final) states of the three-body collisions are labeled by $i$ ($i'$), and the corresponding collision integral involves the generalized Fermi golden rule 
\begin{equation}
W^{123}_{1'2'3'}=\frac{2\pi}{\hbar}|\langle1'2'3'|VG_{0}V|123\rangle_{c}|^{2}\delta(E-E').
\label{Fermi}
\end{equation}
Here, $G_0$ denotes the free Green's function and the subscript $c$ stands for ``connected'' scattering events, in which all three particles participate \cite{Lunde07}. The (unscreened) Coulomb interaction takes the form $V=({1}/{2L})\sum_{k_{1}k_{2}q\sigma_{1}\sigma_{2}}V_{q}a_{k_{1}+q \sigma_{1}}^{\dagger} a_{k_{2} -q\sigma_{2}}^{\dagger}a_{k_{2}\sigma_{2}}a_{k_{1}\sigma_{1}}$ with $V_q\simeq(2e^2/\kappa)\ln(1/qa)$. Our results are dominated by processes with large momentum transfers of order $2k_F$, where screening due to nearby gates can be neglected.

Setting $\delta n_1=\delta_{k_1,k_F+\epsilon/\hbar v_F}$ and neglecting secondary collisions, we obtain, via $(\partial_{t}+\tau^{-1})\delta n_{1}=0$, the total scattering rate of hot particles,
\begin{equation}
\frac{1}{\tau}=\sum_{231'2'3'}W^{123}_{1'2'3'}n_{2}^{0}n_{3}^{0}\bar n_{1'}^{0}\bar n_{2'}^{0}\bar n_{3'}^{0}.
\label{TotalScatRate}
\end{equation}
The transition amplitude for three-body scattering from $|123\rangle$ to $|1'2'3'\rangle$ is the sum of two processes with small momentum transfers $q\ll k_F$ and four processes involving momentum transfers of order $2k_F$, cf.\ Fig.\ \ref{fig:processes}(a)-(c). 

The matrix element in Eq.\ (\ref{Fermi}) can be decomposed as 
\begin{eqnarray}
&&\langle1'2'3'|VG_{0}V|123\rangle_{c} \nonumber\\
&&\,\,\,\,\,\,\,\,\,= \sum_{{P}(1'2'3')}(-1)^{p} \delta_{\sigma_{1}\sigma_{2}\sigma_{3},{P}(\sigma_{1'}\sigma_{2'}\sigma_{3'})}T^{123}_{{P}(1'2'3')}, 
\end{eqnarray}
where $(-1)^p$
denotes the parity of the permutation $P$. Following Ref.\ \cite{Lunde07}, we find for the two small-$q$ processes [cf.\ Fig.\ \ref{fig:processes}(a)], to leading order in $\epsilon/\epsilon_F$,
\begin{equation}
T^{123}_{1'2'3'}  =  (V_{q_{1}}/{8\epsilon_{F}L^2})\left(V_{q_{1}}-V_{q_{3}}\right)
\label{smallq}
\end{equation}
and the corresponding amplitude with $1'\leftrightarrow2'$ (and thus $q_{1}\leftrightarrow p_{1}=k_{1}-k_{2'}$, where $q_{i}=k_{i'}-k_{i}$). Similarly, we find for the four $2k_F$-processes [cf.\ Fig.\ \ref{fig:processes}(b,c)]
\begin{equation}
T^{123}_{3'1'2'}  = -T^{123}_{2'3'1'} = ({V_{2k_{F}}}/{2\hbar v_{F}q_{1}L^2})\left(V_{2k_{F}}-V_{p_{1}}\right)
\label{2kf}
\end{equation}
and the corresponding amplitudes with $1'\leftrightarrow2'$.

Remarkably, the electron spin plays an important role in the transition matrix element for three-body scattering. Indeed, the amplitude of the individual $2k_F$ processes in Eq.\ (\ref{2kf}) is larger than the amplitude of the small-$q$ processes in Eq.\ (\ref{smallq}) by a factor $\epsilon_F/\hbar v_F q_1$. When the total spin of the three colliding particles is maximal, or when fermions are spinless, the leading contribution to the individual $2k_F$ processes cancels from their sum, resulting in total contributions of the small-$q$ and the $2k_F$ processes which are of the same order in $\hbar v_F q_1/\epsilon_F$ (in line with the classification of interactions for spinless fermions). In contrast, when the total spin of the three colliding particles is $1/2$, there is no cancellation in the sum over the $2k_F$ processes. The latter are thus parametrically larger and dominate in the three-body scattering rates. (Note that for Coulomb interactions, the enhancement of order $(\epsilon_F/\epsilon)$ is larger than the concurrent logarithmic reduction stemming from the replacement of $V_{q_1}$ by the smaller $V_{2k_F}$.)

The importance of the electron spin can be understood in terms of the symmetry of the wavefunction. When the total spin of the three colliding particles is maximal, or when fermions are spinless, the orbital part of their wave function must be odd, and the relevant amplitudes are suppressed by the exchange effect. In contrast, no such suppression occurs when the total spin of the three colliding particles is $1/2$, since the interacting particles can be in the same orbital state. 

By integrating over the momenta of the left-moving particle-hole pair and performing the spin sums in Eq.\ (\ref{TotalScatRate}), we can now extract the partial scattering rate $\mathcal{P}\mathrm{d}q_{1}\mathrm{d}p_{1}$ that the hot particle generates co-moving particles near $k_{1'}$ and $k_{2'}$,
\begin{widetext}
\begin{equation}
\mathcal{P}(q_{1},p_{1})=\frac{3\epsilon_{F}}{8\pi^{3}\hbar}\frac{L^4}{(\hbar v_F)^2} 
\frac{q_{3}/k_{F}}{\left({\rm e}^{{\hbar v_Fq_{3}}/{T}}-1\right)}n^0_{k_1+p_1+q_1}\bar n^0_{k_1+p_1} \bar n^0_{k_1+q_1}
{\left[\left(T^{123}_{3'1'2'}\right)^2+\left(T^{123}_{3'2'1'}\right)^2\right]}\label{eq:pel}
\end{equation}
\end{widetext}
By exchange symmetry, the variables $q_1$ and $p_1$ enter symmetrically into Eq.\ (\ref{eq:pel}). In this expression, the factor involving $q_{3}=-{p_{1}q_{1}}/{2k_{F}}$ quantifies the phase space $\sim{\rm max}\{\hbar v_F|q_3|,T\}$ of the left-moving particle-hole pair. When combined with the familiar phase space $\sim\epsilon^2$ of the pair of colliding right-movers, Eq.\ (\ref{eq:pel}) gives a total phase-space factor $\sim \epsilon^2{\rm max}\{T,\epsilon^2/\epsilon_F\}$ for the three-body collision.  

We can now employ Eq.\ (\ref{eq:pel}) to compute the energy relaxation rate of hot particles. Equation (\ref{eq:pel}) shows that the typical energy loss per three-body collision is of order $\epsilon$ so that the energy relaxation rate follows directly from integrating Eq.\ (\ref{eq:pel}) over $q_{1}$ and $p_{1}$. At $T=0$, this integration yields the result given in Eq.\ (\ref{ElRelRate}) above. The dominant $\sim \epsilon^2$ dependence emerges from the total phase space $\sim \epsilon^4$ combined with the singular dependence of the amplitude in Eq.\ (\ref{2kf}) at small $q_1,p_1$, yielding a factor $\sim 1/\epsilon^2$ in the partial scattering rate. 

Once $\epsilon\ll\epsilon_T$ at higher temperatures, the phase space of the left movers is controlled by temperature. In this limit, the total scattering rate diverges logarithmically in the infrared, as $1/\tau\sim\int_{-\epsilon/\hbar v_F}^{0}\mathrm{d}q_{1}/q_{1}$. However, this singularity is regularized in the energy relaxation rate which can be estimated from $1/\tau_{\rm p} = \int dq_1 dp_1(\Delta\epsilon/\epsilon) {\cal P}(q_1,p_1)$ in terms of the energy loss $\Delta\epsilon=\hbar v_F{\rm min} \{|q_1|,|p_1|\}$. This yields the result given above in Eq.\ (\ref{EHRel}). The basic dependences on $\epsilon$ and $T$ can again be understood from the phase-space and amplitude factors. We note in passing that the singularity also does not carry over into the solution of the full Boltzmann equation (\ref{eq:boltzmann}) as long as the injected electron distribution has a finite spectral width. 

The derivation of the hole relaxation rate proceeds in close analogy. A hole injected at $k_{1'}$ generates two holes at $k_1=k_{1'}-q_1$ and $k_2=k_{1'}-p_1$ in a three-body collision. The corresponding partial rate ${\cal P}(q_1,p_1)$ is given by Eq.\ (\ref{eq:pel}) with the replacement $k_1\to k_{1'}$, changes of sign of $q_1$, $p_1$, and $q_3$, and the exchange $n^0 \leftrightarrow \bar n^0$. A crucial modification is the sign change of $q_3$, which limits the small momentum transfer process to $|q_{3}|\lesssim T/\hbar v_F$ and thus the energy loss $\Delta\epsilon$ of holes to $\Delta\epsilon\sim\epsilon_T^2/\epsilon$ when $\epsilon\gg {\epsilon_T}$. Hence, complete energy relaxation proceeds by multiple collisions and the energy relaxation rate $1/\tau_{\rm h}$ can be obtained from $d\epsilon/dt = \int dq_1 dp_1 \Delta\epsilon {\cal P}(q_1,p_1)$ through $\tau_{\rm h} = \int_0^{\epsilon} d\epsilon' (d\epsilon'/dt)^{-1}$. Performing the remaining integrals gives Eq.\ (\ref{HoRel}). This is slower than the relaxation of hot particles by $(\epsilon_T/\epsilon)^4$ since holes not only relax in $(\epsilon/\epsilon_T)^2$ steps, but phase space is also smaller by $T/(\epsilon^2/\epsilon_F)$. As discussed above, the energy relaxation rate becomes equal for particles and holes when $\epsilon\ll\epsilon_T$. 

We briefly comment on the competing process shown in Fig.\ \ref{fig:processes}(d), which involves co-moving electrons only. In this process, two electrons near the Fermi energy are scattered in opposite directions in energy, allowing a high-energy particle to relax slightly. 
Assuming that the interaction is screened for these small momentum transfers, a similar calculation yields their contribution ${1}/{\tau_{\rm p}}\sim ({\epsilon_{F}}/{\hbar}) ({e^{2}}/{\kappa\hbar v_{F}})^{4}({T}/{\epsilon_{{\rm F}}}) ({{\epsilon_T^6}}/ {\epsilon^2\epsilon_0^4}) \ln^4\left|{d}/{a}\right|$ to the energy relaxation rate. (Here, $\epsilon_{0}={\hbar v_{F}}/{d}$.) By comparing with Eq.\ (\ref{ElRelRate}),(\ref{EHRel}), we conclude that these processes are subdominant when $\epsilon\gg\epsilon_T$ or $T\ll \epsilon_F/(k_Fd)^4$ \cite{chiral}.

Finally, we note that the thermalization rate for a given temperature difference $\Delta T$ between right- and left-moving electrons can be obtained in a standard manner by expanding the Boltzmann equation to linear order in $\Delta T$. This yields the result quoted in Eq.\ (\ref{TRel}) above. 

{\em Comparison with experiment.---}The experiment of Ref.\ \cite{Barak10} did not measure the dependence of the energy relaxation time on $\epsilon$, but did provide bounds. Specifically, the energy relaxation rate was inferred to be at least (smaller than) $10^{11}{\mathrm s}^{-1}$ for particles (holes) with $\epsilon$ of order $\epsilon_F/3$. This is consistent with our results which predict \cite{numbers} ${1}/{\tau_{\mathrm{p}}}  \approx  10^{11}{\mathrm{s}}^{-1}$ and ${1}/{\tau_{\mathrm{h}}}  \approx  5\cdot10^{9}{\mathrm{s}}^{-1}$ for $\epsilon\approx\epsilon_F/3$. We also find 
${1}/{\tau_{{\rm inter}}}  \approx  10^{6}{\mathrm{s}}^{-1}$, implying that temperature differences between left- and right-movers can indeed be sustained over long distances.

{\em Conclusion.---}The nonequilibrium physics of one-dimensional electron systems was brought into focus by recent experiments in a variety of contexts such as carbon nanotubes \cite{Chen09}, edge states in quantum Hall systems \cite{Granger09,Altimiras09}, or quantum wires \cite{Auslaender05,Barak10}. Here, we have discussed equilibration and thermalization of hot carriers in real quantum wires, emphasizing the important roles of spin and finite temperature within a perturbative approach for large excitation energies. Our work was motivated by and provides a quantitative framework for a recent experiment \cite{Barak10}. It elucidates a fundamental property of one-dimensional electron liquids and thus has ramifications for one-dimensional electron systems beyond quantum wires. Finally, our work also points to open issues. First and foremost, it would be extremely interesting to understand the fate of energy relaxation in the limit of small excitation energies, where one must simultaneously cope with nonequilibrium physics and the non-perturbative effects of the interactions.

We acknowledge discussions with G.\ Barak, A.\ Imambekov, A.\ Levchenko, T.\ Micklitz, G.\ Refael, and A.\ Yacoby and are grateful for financial support through DFG SPP 1243, DIP, and DOE contract DE-FG02-08ER46482.


\begin{thebibliography}{12}
\bibitem{chiral} Note, however, that processes as in Fig.\ \ref{fig:processes}(d) may dominate energy relaxation in chiral one-dimensional electron systems such as quantum Hall edge states.
\bibitem{numbers} We use $v_{F}=2\cdot10^{5}{{\rm m}}/{{\rm s}}$, $k_{F}=10^{8}{{\rm m}}^{-1}$,
$a\sim20{\rm nm}$, $\kappa=12.4$, and $\epsilon_{F}=6.6{\rm meV}$. Note that for $T=0.25{\rm K}$, $\epsilon_{T}\simeq0.38{\rm meV}$, so that $\epsilon\gg\epsilon_T$. Our estimates use a more accurate $V_q$ than the one after Eq.\ (\ref{Fermi}), valid for $k_Fa\sim 1$.
\end{thebibliography}
\end{document}